\title{Yet another approach to the Maximum Flow}
\author{
        Bj\"orn Hlava \\
                Technische Universit\"at Dortmund\\
	\{bjoern.hlava\}@tu-dortmund.de
}
\date{\today}

\documentclass[12pt]{article}
\usepackage{amsfonts}   
\usepackage{amssymb}
\usepackage{amsmath}
\usepackage[latin1]{inputenc}
\usepackage[T1]{fontenc}
\usepackage{lmodern}
\usepackage{pdfpages}
\usepackage{graphicx}
\usepackage{float}

\usepackage[plain,boxed]{algorithm}
\usepackage{algorithmicx}
\usepackage{algpseudocode}

\begin{document}
\maketitle

\begin{abstract}
I introduce a new approach to the maximum flow problem by a simple algorithm with a slightly better runtime. This approach is based on sorting arcs insight of vertices  on a residual graph. This new approach leads to an $O(mn^{0.5})$ time bound for a network with $n$ vertices and $m$ arcs.\\
The Category: Algorithms, Graph Theory and maximum flows
\end{abstract}

\section{Introduction}
The maximum flow problem and the minimum cut problem have a big variety of uses in scientific and engineering applications. Even if it's one of the well understood combinatorial problems with a long history and many approaches to improve the asymptotic runtime.\\
In this paper an algorithm which combines calculation of blocking flows, shortest path and the usage of preflow\cite{prf}, will be introduced. By intensively studying the hi\_prf the author wanted to develop a non numerical topology based algorithm to get rid of the relabel function. Coming up with the idea, that it might be possible to know where the excess needs to move, can be calculated faster. For now that can be done, but not for a single vertex. More for using something similar to a global update\cite{iprf} to calculate something similar to an acyclic core, known from Binary Blocking Flow \cite{bbf}.\\
The advantage is the speed that core can be calculated in $O(m)$, even though that calculation is very simple it generates all necessary informations, to know all dependencies for each residual network, for minimizing the amount of non saturating pushes. Though there is only a global update, which is using unit distance.\\
During the global push all improving dependencies will be augmented. As shown later this will also only need $O(m)$, because every arc of the core will be only once augmented.\\
Even with worst case improvement of each iteration it can be shown, that those two functions only need to be called $O(n^{0.5})$.

\section{Definitions and Notation}
The input to the maximum flow problem is a network $N(G(V,E),s,t,u)$, where $s,t \in V : s\neq t$ are source and sink and $u: A \rightarrow \mathbb{N}$ is the capacity function. Let $n=|V|$ and $m=|A|$.\\
A flow in a network is a function $f: E \rightarrow \mathbb{N}$ where for each arc, the capacity constraints $0 \leq f(a) \leq u(a)$ holds and for each vertex $j \in V \not \quad \{s, t\}$ the conservation constraint $\sum_{(j,k)}f(j,k) = \sum_{(i,j)} f(i,j) = 0$ holds. The value of the flow is $|f|=\sum{(j,t)} f(j,t)$. $f(a)$ is referred as the flow on arc $a$.\\
A preflow is relaxing the conservation constraint to be $\sum_{(j,k)}f(j,k) \leq \sum_{(i,j)} f(i,j)$ resulting in an excess $e(j) = \sum_{(i,j)} f(i,j) - \sum_{(j,k)}f(j,k)$.
We assume that every arc $a$ has a reverse arc $a^R$. The residual capacity $u_f$ of an arc $(i,j)$ is defined to be $u(i,j) -  f(i,j) + f(j,i)$.\\
Let $o$ be a function $\forall v \exists o:\mathbb{N}\rightarrow (v,w) \in A$ order of outgoing arcs for each vertex.

\section{Sorting Flow}
The algorithm has only two operations effecting the whole graph. The first we call Breadth First Search Sort or short BFSS, which calculates the arcs which need to be augmented. The second is a global push operation augmenting these arcs. These two operations are repeated until no augmenting path can be found. BFSS will return true, when an augmenting path exists.
\begin{algorithm}[H]
\caption{Sorting Flow}
\label{sf}
\begin{algorithmic}[1]
\Require{$N(G(V,E),s,t,u)$}
\Ensure{$|f|$ to be maximized}
\While {BFSS}
\State Push()
\EndWhile
\end{algorithmic}
\end{algorithm}
\subsection{BFSS}
This function has two objectives. First: checking for an augmenting path, which here means $\overline{vt} : e(v)>0$, where $\forall a \in \overline{vt} : u_f(a)>0$. The second objective is to sort all arcs in every vertex by their shortest distance. To achieve this we run a regular BFS starting with $t$, using a queue and looking at ingoing arcs. During this every vertex has one of three states: 0 not found, 1 found and 2 used. Every vertex in state 1 is element of the queue.\\
In initial all nodes are state 0, except $t$, which also is in the queue. While the que isn't empty the algorithm removes the first element of the queue, marks it as state 2 and now iterates through the ingoing arcs $(u,v)$. For all vertices $u$ which state is 0, set their state to 1 and queue them to the queue. For all vertices with a state $\neq$ 2 set the next free value in $o(u)$ to the arc $(u,v)$.\\
When ever a vertex $v$ is removed from that queue, if $e(v)>0$ or $v = s$, put this node into another queue. to not mess up those to queues the second one will be referred as priority queue.\\
At the end of this function we have a priority queue, with all vertices $e_f(v) > 0$. Now looking only on arcs addressed by function $o$ reveals an acyclic core. The ordering insight of the priority queue matches for every residual graph with with acyclic core.
\begin{algorithm}[H]
\caption{Breadth First Search Sort}
\label{bfss}
\begin{algorithmic}[1]
\State final = true;
\State que.push(t)
\While {!que.empty()}
\State $v$ = que.getFront()
\If {$e(v)>0$}
\State priority queue.pushLast(v)
\EndIf
\State $v$.state=2 
\ForAll{$u_f(u,v) >0 \in E$}
\State add $(u,v)$ to $o(u)$
\If {$u$.state $\neq 2$}
\State que.pushLast($u$)
\State $u$.state=1
\EndIf
\EndFor
\EndWhile
\end{algorithmic}
\end{algorithm}

\subsection{Push}
In this step the excess will be pushed as much as possible. By using the priority queue and only augmenting arcs, which are addressed by the function $o$. As long as the priority queue isn't empty the algorithm will remove the first node $v$ and will as long as there are arcs in $o(v)$ and excess $e(v)>0$. All arcs of $o(v)$ will be augmented. If the the excess of another vertices changes from $0$ to $>0$, those vertices will be put to the front of the priority queue to be executed next.
\begin{algorithm}[H]
\caption{Push}
\label{gp}
\begin{algorithmic}[1]
\While {!priority queue.empty()}
\State $v$=priority queue.getFront();
\ForAll{$(v,w) \in o(v)$ or until $e_f(v)=0$}
\If{$e_f(w)=0$}
\State priority queue.pushFront($w$)
\EndIf
\State augment $(v,w)$
\EndFor
\EndWhile
\end{algorithmic}
\end{algorithm}

\section{Correctness and Runtime}
\subsection{Correctness}
Even in a worst case situation BFSS will find at least one augmenting path, if this exists. This path will be found in $\forall o(w) \in \overline{vt} : e_f(v) > 0$. The Push will increase the excess of the sink at least by one and saturate one arc or discharge one vertex. If no augmenting path $\overline{vt} : e_f(v)>0$ can be found the algorithm determines. According to Ford and Fulkerson this blocking flow is supposed to be maximal.
\subsection{Runtime}
\begin{figure}[h]
\center
\includegraphics[scale=0.5]{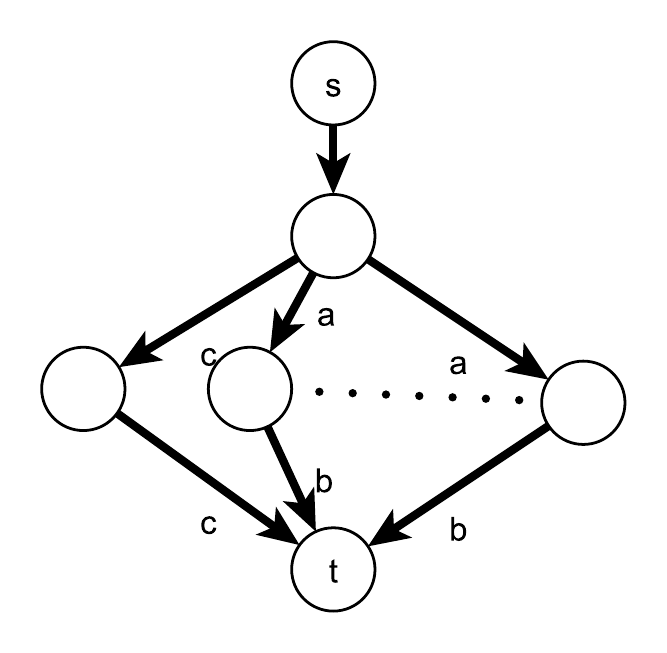}
\includegraphics[scale=0.5]{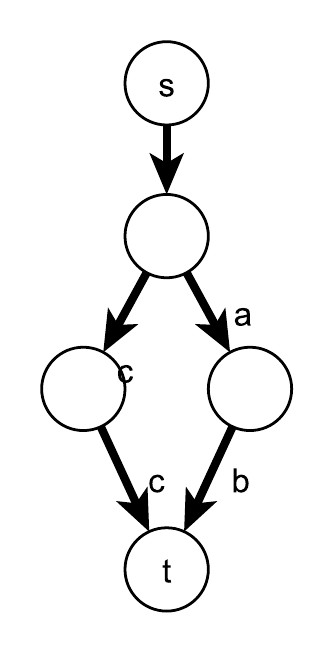}
\caption{to be solved in two iterations}
\label{2it}
\end{figure}

\begin{figure}[h]
\center
\includegraphics[scale=0.5]{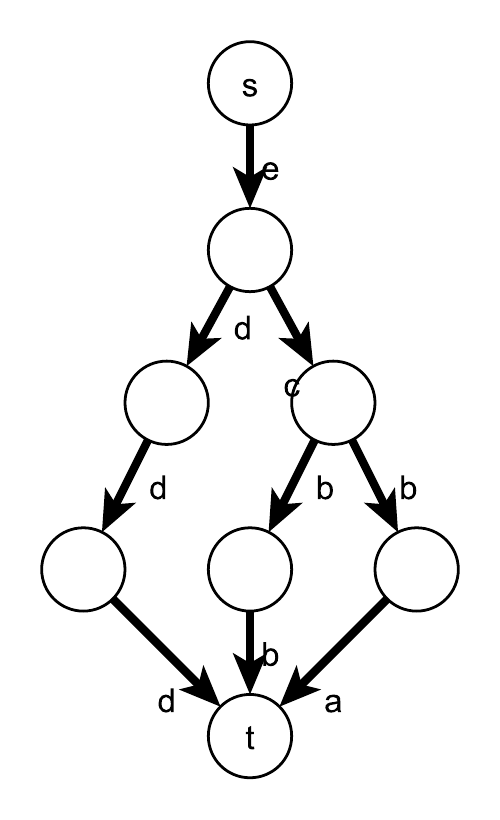}
\includegraphics[scale=0.5]{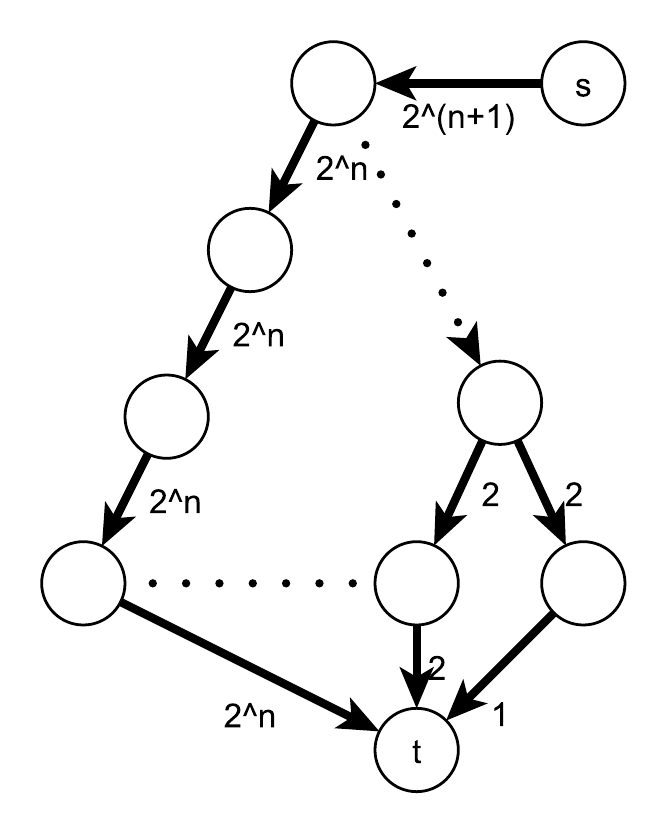}
\caption{to be solved in three and $n-1$ iterations}
\label{3it}
\end{figure}

BFSS and Push look at every arc only once, therefor the runtime for each call of these functions has to be $O(m)$.\\
\begin{figure}[h]
\center
\includegraphics[scale=0.5]{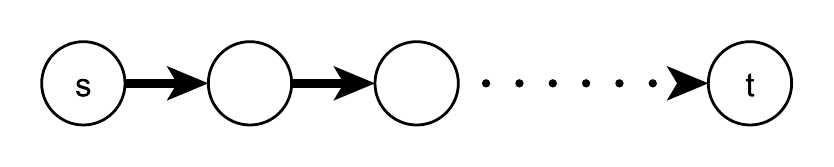}
\includegraphics[scale=0.5]{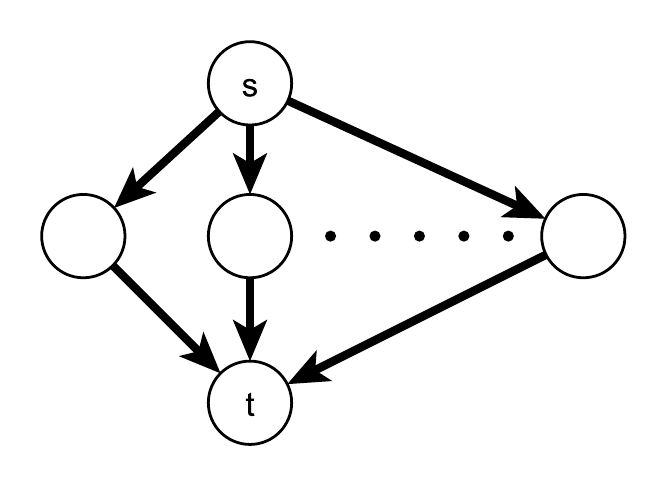}
\caption{to be solved in one iteration}
\label{1it}
\end{figure}
The amount of calls for those two functions can be easily understood, when thinking about the smallest worst case inputs. Starting with the graphs, which can be solved in one iteration. Those can be described as a simple line as seen in figure \ref{1it}. To force a second iteration it is necessary to block the flow at one arc, as seen in figure \ref{2it}. If $d > c > b > a$ the excess will stop on the middle layer of this graph. After the next iteration is done. The max flow will be calculated.\\
This can be done for even one layer more or an unlimited amount of layer as seen in figure \ref{3it}. When deforming these graphs, and observing the algorithm in their worst case it still can be observed, that the algorithm completes one layer breadth and one layer length in each iteration. concluding to an amount of calls $O(min(length(f), breadth(f) ) \subseteq O(n^{0.5})$, where $f$ is the calculated maximum blocking flow.\\

\nocite{prf}

\bibliographystyle{abbrv}
\bibliography{main}

\end{document}